# Harvest Season for SLUB:

## From io_uring vulnerability to Novel Sheaf-Based Exploitation Techniques

Hao-Yu Yang(naup96321)[1] and Yu-Ting Lin[2]

[1]naup96721@gmail.com
[2]dong1214.mailbox@gmail.com

August 13, 2026

**Abstract**

The Linux kernel's push for higher I/O performance and more efficient memory management has introduced new mechanisms that, while improving performance, also open new attack surfaces. This research examines two of them together: the io_uring subsystem and the sheaf/barn caching mechanism added to the SLUB allocator in Linux 6.18.

In this research, two previously unknown vulnerabilities in io_uring are presented, and one is developed into a complete local privilege escalation chain under a hardened kernel configuration. Building this chain revealed that the sheaf/barn mechanism changes long-standing assumptions behind established exploitation techniques such as cross-cache attack, and that its design also weakens existing SLUB freelist protections. Both observations are analyzed and turned into working primitives.

Building on this analysis, three novel sheaf-based exploitation techniques are proposed. Among them, an RCU-sheaf cross-cache technique removes the traditional dependence on the buddy system for moving objects across caches, giving more flexible and reliable control over object migration between cache pools. Together, these results characterize the sheaf/barn layer as a new and largely unexplored attack surface in Linux kernel exploitation.

# Contents

# 1 Background

## 1.1 io_uring Overview

io_uring is a kernel core component for high-performance asynchronous I/O operations, first introduced in Linux kernel 5.1 [1]. It primarily addresses two problems: the performance bottleneck caused by synchronous blocking in traditional I/O operations, and the cost of numerous system calls. io_uring also provides a broad async I/O interface, covering file systems, sockets, polling, and more.

As shown in Figure 1, io_uring achieves this by constructing a Submission Queue (SQ) and a Completion Queue (CQ) in shared memory between the kernel and user space. An application in user space packs each required I/O operation into a Submission Queue Entry (SQE) and places it into the SQ. Then the kernel reads the SQE and performs the corresponding operations. Once the operation is complete, the kernel writes the result to the Completion Queue Entry (CQE), which is then read by the application in user space. This design improves the performance by avoiding a system call for most I/O operations.

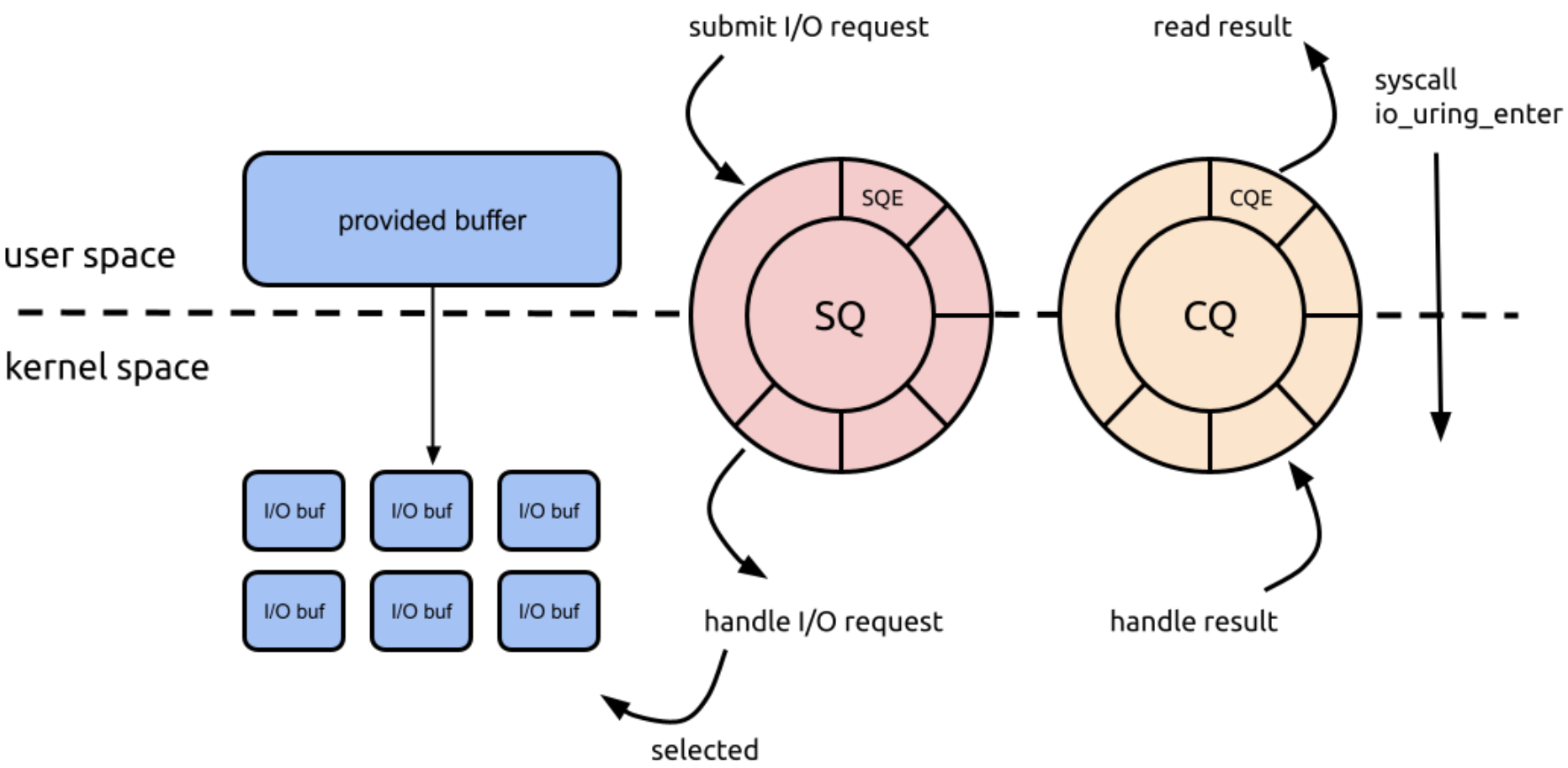

Figure 1: io_uring submission and completion queues shared between user space and the kernel.

In addition, io_uring encourages processing multiple SQEs at once. An application can submit multiple requests to the SQ in a single batch, and the kernel can likewise fill multiple completion events into the CQ at once. There are two ways to trigger the kernel to process the I/O requests in the SQ:

- The `io_uring_enter` system call notifies the kernel to process the I/O requests currently in the SQ.
- With `SQPOLL`, the kernel starts an `iou-sqp` thread to poll the SQ, so that some I/O operations can be submitted to the kernel without the application issuing another system call, further improving efficiency.

Unlike traditional `epoll`, which mainly notifies an application whether a file descriptor is ready for an operation, the actual I/O still has to be carried out by the application itself. In comparison, io_uring lets an application submit a complete I/O request and has the kernel perform the operation and report the result once it is done. This reduces the complexity of managing the I/O flow on the application side. It also resolves several performance problems found in AIO. AIO supports only a limited range of I/O types, its interface design is relatively fragmented, and each I/O request usually requires a dedicated control structure along with multiple system calls for submission and status queries [2, 3].

## 1.2 SLUB and the Sheaf Mechanism

As shown in Figure 2, the SLUB allocator is the kernel component responsible for managing memory allocation and freeing. Beneath it, many different `kmem_cache` form memory pools, each responsible for managing the slabs of a specific data structure type or a fixed memory block size. A slab is a set of one or more contiguous pages of memory, and these pages hold kernel objects of a specific size.

In traditional SLUB design, a `kmem_cache` is managed mainly through two levels: a CPU-local cache layer (`kmem_cache_cpu`) and a NUMA node layer (`kmem_cache_node`). As core counts grew and large NUMA systems became common, the traditional per-CPU cache could still incur synchronization and management costs when many CPUs operated simultaneously. Linux 6.18 [4] therefore introduces a new SLUB object caching mechanism that gradually replaces the local object management previously handled by `kmem_cache_cpu`, building a new caching layer out of sheaf and barn [5]. The current SLUB architecture can thus be viewed as three layers that together manage kernel objects: the sheaf layer, the barn layer, and the slab management layer (`kmem_cache_node`).

This layer is composed of three `slab_sheaf` slots: `main`, `spare`, and `rcu_free`. `main` is the sheaf currently in use, and any object to be allocated or freed is taken from this layer's `main` slot. `spare` serves as a backup and is used as a replacement when the `main` slot has no object left to hand out or no room to store one. `rcu_free` is used for batch processing of RCU objects after `kfree_rcu`.

mm/slub.c

```c
// https://elixir.bootlin.com/linux/v7.1.4/source/mm/slub.c#L420
struct slub_percpu_sheaves {
  local_trylock_t lock;
  struct slab_sheaf *main; /* never NULL when unlocked */
  struct slab_sheaf *spare; /* empty or full, may be NULL */
  struct slab_sheaf *rcu_free; /* for batching kfree_rcu() */
};
```

### 1.2.1 Layer 2: the barn layer (`node_barn`)

This layer keeps completely empty or completely full sheaves as reserves, which are shared within the same NUMA node. When the per-CPU sheaves are exhausted, a spinlock is acquired, and the sheaves in the barn are swapped with those in the per-CPU layer. The upper limits on the sheaves stored in the barn are `MAX_FULL_SHEAVES` and `MAX_EMPTY_SHEAVES` respectively, constants that will be referenced again during exploitation.

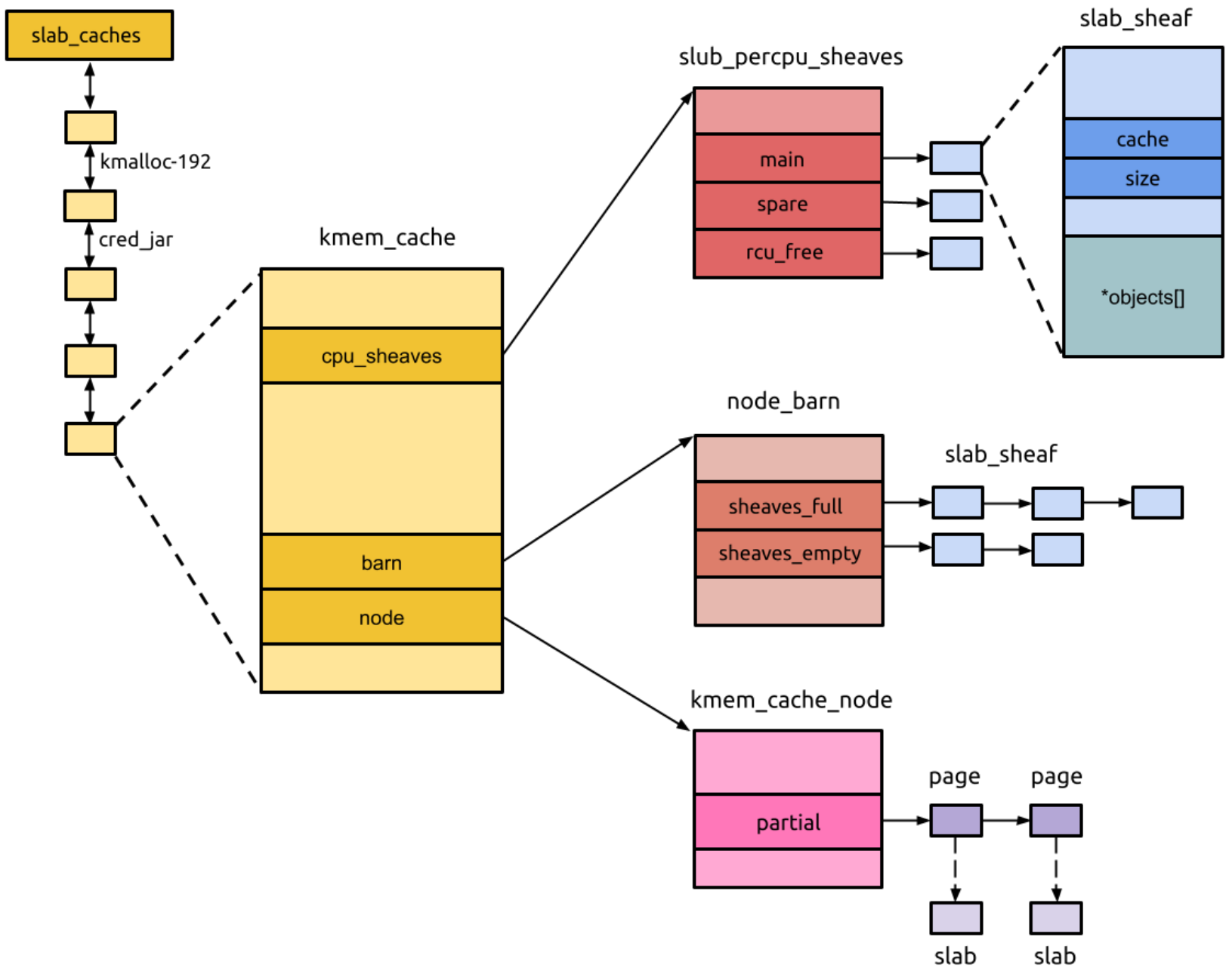

Figure 2: The three-layer SLUB object management architecture after the introduction of sheaves.

mm/slub.c

```
// https://elixir.bootlin.com/linux/v7.1.4/source/mm/slub.c#L396
struct node_barn {
  spinlock_t lock;
  struct list_head sheaves_full;
  struct list_head sheaves_empty;
  unsigned int nr_full;
  unsigned int nr_empty;
};
```

### 1.2.2 Layer 3: the slab management layer (`kmem_cache_node`)

`kmem_cache_node`, the NUMA node layer, manages slab state at the NUMA node level, handles memory reclamation and balancing across CPUs or nodes, and maintains the complete list of all slabs on that node, including partial, full, and empty slabs.

# 2 Vulnerability 1: Race Condition Between CQ/SQ Resize and Deferred Task Work

## 2.1 Root Cause

The vulnerability can be triggered through the `io_uring_register` syscall. When the opcode is set to `IORING_REGISTER_RESIZE_RINGS`, the request is forwarded to `io_register_resize_rings()`.

io_uring/register.c

```
// https://elixir.bootlin.com/linux/v6.19.3/source/io_uring/register.c#L805
case IORING_REGISTER_RESIZE_RINGS:
    ret = -EINVAL;
    if (!arg || nr_args != 1)
        break;
    ret = io_register_resize_rings(ctx, arg);
    break;
```

`IORING_REGISTER_RESIZE_RINGS` allows a process to expand or shrink the SQ/CQ ring of an existing io_uring instance without recreating the entire io_uring. `io_register_resize_rings()` first acquires `ctx->mmap_lock` and `ctx->completion_lock`, then temporarily sets the rings pointed to by the `io_ring_ctx` of the I/O request to NULL. After the new ring memory is allocated, the newly allocated address is written back into the `io_ring_ctx`.

io_uring/register.c

```
// https://elixir.bootlin.com/linux/v6.19.3/source/io_uring/register.c#L489
  mutex_lock(&ctx->mmap_lock);
  spin_lock(&ctx->completion_lock);
  o.rings = ctx->rings;
  ctx->rings = NULL;
  o.sq_sqes = ctx->sq_sqes;
  ctx->sq_sqes = NULL;
  ...
  ctx->sq_entries = p->sq_entries;
  ctx->cq_entries = p->cq_entries;

  ctx->rings = n.rings;
```

Another component that interacts with this feature and causes the vulnerability is `IORING_SETUP_DEFER_TASKRUN`. As described in the official documentation [6], io_uring by default processes all outstanding work at the end of any system call or thread interrupt, which can delay the application's further progress. Setting this flag will hint to io_uring that it should defer work until an `io_uring_enter` call with the `IORING_ENTER_GETEVENTS` flag set.

These deferred tasks are first placed into the task work queue. When new task work is added, the kernel needs to set `IORING_SQ_TASKRUN` to notify userspace that there are pending tasks that need to be processed. However, when updating the corresponding flags in `io_ring_ctx`, the operation does not acquire the same lock used by the resize functionality, resulting in a race condition.

io_uring/io_uring.c

```
// https://elixir.bootlin.com/linux/v6.19.3/source/io_uring/io_uring.c#L1258
static void io_req_local_work_add(struct io_kiocb *req, unsigned flags)
{
  struct io_ring_ctx *ctx = req->ctx;
  ...
  if (!head) {
    if (ctx->flags & IORING_SETUP_TASKRUN_FLAG)
      atomic_or(IORING_SQ_TASKRUN, &ctx->rings->sq_flags);
    if (ctx->has_evfd)
      io_eventfd_signal(ctx, false);
  }
}
```

## 2.2 Proof of Concept

First, create an io_uring instance using `IORING_SETUP_DEFER_TASKRUN` and mmap the io_uring ring into userspace. Then create a `timerfd` that triggers at 100-microsecond intervals, and submit a multishot `IORING_OP_POLL_ADD` request to monitor the `timerfd` for the `POLLIN` event. Each time the timer expires, the kernel generates a poll completion. In `DEFER_TASKRUN` mode, these completions are deferred through task work, causing the kernel to repeatedly execute `io_req_local_work_add()` and set the `IORING_SQ_TASKRUN` flag.

In the main loop that triggers the race condition, each iteration first invokes `io_uring_enter(GETEVENTS)`, allowing the kernel to execute the currently accumulated task work. It then calls `IORING_REGISTER_RESIZE_RINGS` to reconfigure the current SQ/CQ rings. If, at some point, the task work is about to modify `sq_flags` through `ctx->rings` while another CPU is concurrently performing a ring resize operation, or if the execution flow is preempted by a ring resize task during the critical timing window, the race condition may cause the ring structure referenced by `ctx->rings` to become invalid, ultimately resulting in a null pointer dereference.

## Disclosure Timeline

- 2026-03-07: Vulnerability discovered
- 2026-03-09: Reported upstream together with a proposed patch
- 2026-03-17: Patch merged and released

## Artifacts

- Patch: https://git.kernel.org/pub/scm/linux/kernel/git/stable/stable-queue.git/diff/queue-6.18/io_uring-ensure-ctx-rings-is-stable-for-task-work-flags-manipulation.patch?id=511c1f4e8933744a6e7475c5defaad673093215b
- Exploit: https://hackmd.io/epHcDaB_Tr6RZWmLobJY0g?view#vulnerability-1-PoC
- Demo: https://youtu.be/-vPd8k4ix4g

# 3 Vulnerability 2: Logic Error in the Error-Handling Path of the Provided-Buffer Bundle Expansion Flow

## 3.1 Root Cause

The vulnerability appears on the handling path of `IORING_OP_SEND` [7], which is the asynchronous socket send operation provided by io_uring for submitting a socket data transmission request. When a user submits this operation through io_uring, the kernel enters the `io_send` function to process the request.

When the request adopts the buffer selection mechanism, the handling flow enters the buffer selection path of send. At this point, the kernel builds the `buf_sel_arg` and selects a suitable buffer from the registered buffer pool through `io_buffers_select()` to serve as the data source for the subsequent send operation.

io_uring/net.c

```
// https://elixir.bootlin.com/linux/v7.2-rc1/source/io_uring/net.c#L620
    struct buf_sel_arg arg = {
        .iovs = &kmsg->fast_iov,
        .max_len = min_not_zero(sr->len, INT_MAX),
        .nr_iovs = 1,
        .buf_group = sr->buf_group,
    };
    int ret;

    if (kmsg->vec.iovec) {
        arg.nr_iovs = kmsg->vec.nr;
        arg.iovs = kmsg->vec.iovec;
        arg.mode = KBUF_MODE_FREE;
    }

    if (!(sr->flags & IORING_RECVSEND_BUNDLE))
        arg.nr_iovs = 1;
    else
        arg.mode |= KBUF_MODE_EXPAND;

    ret = io_buffers_select(req, &arg, sel, issue_flags);
    if (unlikely(ret < 0))
        return ret;

    if (arg.iovs != &kmsg->fast_iov && arg.iovs != kmsg->vec.iovec) {
        kmsg->vec.nr = ret;
        kmsg->vec.iovec = arg.iovs;
        req->flags |= REQ_F_NEED_CLEANUP;
    }
```

At this time, an `iovec` pointer is copied from `kmsg` into `arg.iovs` and then passed as an argument to `io_buffers_select()`. `io_buffers_select()` is responsible for locating the buffer list based on `buf_group`. For convenience, this research chooses the provided buffer as the further analysis target.

A provided buffer allows an application to perform a receive or send without specifying a concrete memory address in each request. The kernel automatically picks free memory from a pre-registered buffer pool to hold the data, and these buffers can be registered through `IORING_REGISTER_PBUF_RING` [8].

```
io_uring/kbuf.c
// https://elixir.bootlin.com/linux/v7.2-rc1/source/io_uring/kbuf.c#L352
int io_buffers_select(struct io_kiocb *req, struct buf_sel_arg *arg,
          struct io_br_sel *sel, unsigned int issue_flags)
{
  struct io_ring_ctx *ctx = req->ctx;
  int ret = -ENOENT;

  io_ring_submit_lock(ctx, issue_flags);
  sel->buf_list = io_buffer_get_list(ctx, arg->buf_group);
  if (unlikely(!sel->buf_list))
    goto out_unlock;

  if (sel->buf_list->flags & IOBL_BUF_RING) {
    ret = io_ring_buffers_peek(req, arg, sel->buf_list);
```

`io_ring_buffers_peek()` starts from the head of the buffer ring and takes out one or more buffers in order, converting them into `iovec` for subsequent send/recv use. When the bundle feature is enabled, it allows multiple buffers, that is multiple `iovec`, to be selected. At this point `kmsg` holds an `iovec` array, and if that array is not large enough, it is expanded by allocating a new region with `kmalloc`.

```
io_uring/kbuf.c
// https://elixir.bootlin.com/linux/v7.2-rc1/source/io_uring/kbuf.c#L287
    iov = kmalloc_objs(struct iovec, nr_avail);
    if (unlikely(!iov))
      return -ENOMEM;
    if (arg->mode & KBUF_MODE_FREE)
      kfree(arg->iovs);
    arg->iovs = iov;
    nr_iovs = nr_avail;
```

Once processing finishes and returns upward, the updated `iovec` in `arg` is flushed back to `kmsg` so that `kmsg` holds the new `iovec`.

However, after the expansion, when each provided buffer address is checked individually, if an illegal address is found, an error is returned. This error is early-returned before the `kmsg` pointer is updated, so `kmsg` still holds the pointer to the old `iovec` array that has already been freed, resulting in a use-after-free.

```
io_uring/kbuf.c
// https://elixir.bootlin.com/linux/v7.2-rc1/source/io_uring/kbuf.c#L318
    if (unlikely(!access_ok(iov->iov_base, len))) {
      if (arg->iovs != org_iovs)
        kfree(arg->iovs);
      return -EFAULT;
    }
```

## 3.2 Exploitation

In the test environment, the following protections are enabled:

- KASLR, KPTI, SMAP, SMEP

- CONFIG_MEMCG_KMEM
- CONFIG_STATIC_USERMODEHELPER
- CONFIG_SLAB_FREELIST_HARDENED
- CONFIG_SLAB_FREELIST_RANDOM

### 3.2.1 Leaking KASLR

This is a powerful primitive. The UAF structure allows the `kmalloc` size to be controlled, its memory pool is `GFP_KERNEL`, and data can be written onto the UAF structure. However, since `CONFIG_MEMCG_KMEM` is enabled, the target structure should avoid using the `kmalloc` structure with the `GFP_KERNEL_ACCOUNT` flag at `kmalloc` time when heap spraying. An alternative solution is to conduct a cross-cache, an innovative version of which is discussed in Sections 4 and 6 [9].

The procedure for obtaining the UAF afterward follows the flow described below, except that the initial `iovec` size needs to be adjusted according to the size of the structure to be reclaimed.

First, add a key to pre-allocate `user_key_payload`, and create a `socketpair` in advance for later io_uring use. Then create a uring and map the CQ and SQ.

**First sending**

A provided buffer is created and registered to the uring. At this stage, 4 provided buffers are created, with 4 `iovec` coming from kmalloc-64. An SQE is then set up to send the data inside the provided buffer. When `IOSQE_BUFFER_SELECT` [10] is specified, an I/O buffer is selected from the provided buffers as the request for the socket to read or receive data, and `IORING_RECVSEND_BUNDLE` [11] must also be specified here. With this, the send operation tries to fill multiple buffers at once rather than selecting only a single buffer to transmit, which is what makes the `iovec` expansion reachable. Once the SQE is set up, the first send can be issued, and `kmsg` now holds the `iovec` array of 4 entries.

**Second sending**

Next, 8 provided buffers are set up again, at index 4 to 11. One of these provided buffer addresses is made to point to an illegal address, used to trigger an `access_ok` failure, which makes the old `iovec` object held by `kmsg` be freed. After the `access_ok` failure, the entire I/O request fails, and the subsequent cleanup frees the `iovec` again, causing a double free. The reason this does not trigger the `CONFIG_SLAB_FREELIST_HARDENED` double free detection is covered in Section 5.1.

**Third sending**

`update_key` is used to reallocate the key structure, and a third send is issued to make the `iovec` array overlap with the `user_key_payload` structure.

```c
// https://elixir.bootlin.com/linux/v7.2-rc1/source/security/keys/keyctl.c#L325
long keyctl_update_key(key_serial_t id,
           const void __user *_payload,
           size_t plen)
{
  ...
  /* pull the payload in if one was supplied */
  payload = NULL;
  if (plen) {
    ret = -ENOMEM;
    payload = kvmalloc(plen, GFP_KERNEL);
    if (!payload)
      goto error;
```

The `user_key_payload` structure is shown below. `datalen` at offset `0x10` is metadata that controls the length of the key data, and it overlaps the low 2 bytes of the second element's address in the `iovec` array. Therefore, when supplying the provided buffer, an address whose low bits are `0x1000` is mapped in advance to overwrite the key structure.

```c
// https://elixir.bootlin.com/linux/v7.2-rc1/source/include/keys/user-type.h#L27
struct user_key_payload {
  struct rcu_head rcu;     /* RCU destructor */
  unsigned short  datalen;  /* length of this data */
  char    data[] __aligned(__alignof__(u64)); /* actual data */
};
```

Finally, reading the key back produces an out-of-bounds read that leaks KASLR.

### 3.2.2 Leaking the `task_struct` address

The idea for privilege escalation is to leak the `task_struct` and overwrite its `cred` pointer with `init_cred`. Leaking the current process's `task_struct` address relies on `io_ring_ctx`, a kmalloc-2048 sized object. When created with `IORING_SETUP_SINGLE_ISSUER` [12] set, the io_uring instance is bound to the current task and can only be submitted by that single task, and the `task_struct` is recorded in the `io_ring_ctx`.

```c
// https://elixir.bootlin.com/linux/v7.2-rc1/source/io_uring/io_uring.c#L3065
  if (ctx->flags & IORING_SETUP_SINGLE_ISSUER
      && !(ctx->flags & IORING_SETUP_R_DISABLED))
    ctx->submitter_task = get_task_struct(current);
```

By heap spraying this structure, and again using the same procedure as before, a `user_key_payload` capable of an out-of-bounds read is produced. Here, `setxattr` is used to overlap with the key structure from kmalloc-2048.

fs/xattr.c

```
// https://elixir.bootlin.com/linux/v7.2-rc1/source/fs/xattr.c#L645
    ctx->kvalue = vmemdup_user(ctx->cvalue, ctx->size);
    if (IS_ERR(ctx->kvalue)) {
      error = PTR_ERR(ctx->kvalue);
      ctx->kvalue = NULL;
    }
```

In this way, the `task_struct` address of the current process can be leaked.

### 3.2.3 One-shot cred overwrite leading to local privilege escalation

Since the `task_struct` address is leaked, the next step is to write to the `cred` pointer of `task_struct`. The setup is basically the same: a kmalloc-256 double free into the sheaf is performed, and `setxattr` is overlapped with `msg_msg`, which is allocated once `msgsnd` is issued.

To overwrite the `cred` pointer in `task_struct`, the overlapped `msg_msg` is abused for arbitrary writes. Specifically, `m_list.next` is written as `task_struct->real_cred`, `m_list.prev` as `init_cred`, and `security` as a readable address. `security` must be set to a readable address because during `msgrcv` the SELinux hook `security_msg_queue_msgrcv` accesses that field.

include/linux/msg.h, ipc/msgutil.c

```
// https://elixir.bootlin.com/linux/v7.2-rc1/source/include/linux/msg.h#L9
struct msg_msg {
  struct list_head m_list;
  long m_type;
  size_t m_ts;    /* message text size */
  struct msg_msgseg *next;
  void *security;
  /* the actual message follows immediately */
};

// https://elixir.bootlin.com/linux/v7.2-rc1/source/ipc/msgutil.c#L54
static struct msg_msg *alloc_msg(size_t len)
{
  struct msg_msg *msg;
  struct msg_msgseg **pseg;
  size_t alen;

  alen = min(len, DATALEN_MSG);
  msg = kmem_buckets_alloc(msg_buckets, sizeof(*msg) + alen, GFP_KERNEL);
  if (msg == NULL)
    return NULL;
```

When `list_del` is finally triggered, the `cred` pointer is written with `init_cred`, successfully completing the privilege escalation to root, as `task_struct->real_cred + 8` is exactly `task_struct->cred`.

ipc/msg.c, include/linux/list.h

```c
// https://elixir.bootlin.com/linux/v7.2-rc1/source/ipc/msg.c#L1163
// https://elixir.bootlin.com/linux/v7.2-rc1/source/include/linux/list.h#L224
static inline void __list_del(struct list_head * prev, struct list_head * next)
{
  next->prev = prev;
  WRITE_ONCE(prev->next, next);
}
```

### Disclosure Timeline

- 2026-07-03: Vulnerability discovered
- 2026-07-06: Reported upstream together with a proposed patch
- 2026-07-12: Patch merged and released

### Artifacts

- Patch: https://git.kernel.org/pub/scm/linux/kernel/git/axboe/linux.git/commit/?h=io_uring-7.2&id=cd053d788c3f13b3eaf16672d427ee828fda16ed
- Exploit: https://hackmd.io/epHcDaB_Tr6RZWmLobJY0g?view#vulnerability-2-exploit
- Demo: https://youtu.be/xp5Qc041YCY

# 4 The Impact of Sheaf on Traditional Exploitation and the Solutions

As mentioned in Section 3.2.1, the mechanism of the SLUB allocator has changed significantly since Linux kernel 6.18, due to the sheaf/barn cache layer described in Section 1.2, which causes the traditional cross-cache method to fail. This section analyzes the cause, proposes two solutions to the impact of sheaf on prior SLUB techniques, and presents one approach that can bypass the sheaf cache entirely and exploit the old slab layer directly.

## 4.1 The Impact of the Sheaf Mechanism

In the traditional cross-cache techniques[13][14][15], a large number of freed slabs are produced, including the ones holding UAF objects, to exhaust the partial slab space of the corresponding `kmem_cache`. This makes the buddy system reclaim the page so that the page of the UAF object is allocated by another cache, and the UAF structure ends up overlapping with a structure from a different `kmem_cache`.

However, sheaf directly affects the reclamation flow of a slab page. When an object in a slab is cached by sheaf, even though the object has been logically freed, it is still regarded as an object in use from the slab's point of view. This prevents the slab holding the UAF object from entering the expected reclamation state, so the corresponding page cannot be reclaimed by the buddy system and reallocated to another `kmem_cache`.

This is also the main reason why the traditional cross-cache method did not work while analyzing the exploitation flow of the provided buffer vulnerability in this research. Although the release of the UAF object can be triggered, the slab page it sits on can never be reclaimed by the buddy system, so it is not possible to overlay that memory region by allocating objects through another `kmem_cache`, and the intended cross-cache attack cannot succeed.

## 4.2 Solution 1: Flushing Sheaf and Barn Objects Back to the Slab Management Layer

The first solution adds an extra step during cross-cache. The poisoned sheaf is first swapped from the per-CPU layer into the barn, and once the full sheaves or empty sheaves in the barn reach `MAX_FULL_SHEAVES` or `MAX_EMPTY_SHEAVES`, they are flushed back into the slab.

```
mm/slub.c

// https://elixir.bootlin.com/linux/v7.1.4/source/mm/slub.c#L393
#define MAX_FULL_SHEAVES    10
#define MAX_EMPTY_SHEAVES   10

// https://elixir.bootlin.com/linux/v7.1.4/source/mm/slub.c#L5722
  empty = barn_replace_full_sheaf(barn, pcs->main, allow_spin);
  ...
  /* sheaf_flush_unused() doesn't support !allow_spin */
  if (PTR_ERR(empty) == -E2BIG && allow_spin) {
    ...
    sheaf_flush_unused(s, to_flush);
  }
```

`__kmem_cache_free_bulk()` eventually reaches `__slab_free()` and returns the object back into the slab. Only then does `slab->inuse` drop, allowing the buddy system to reclaim the slab's pages.

```
mm/slub.c

// https://elixir.bootlin.com/linux/v7.1.4/source/mm/slub.c#L2914
static void sheaf_flush_unused(struct kmem_cache *s, struct slab_sheaf *sheaf)
{
  ...
  __kmem_cache_free_bulk(s, sheaf->size, &sheaf->objects[0]);
}

// https://elixir.bootlin.com/linux/v7.1.4/source/mm/slub.c#L7063
static void __kmem_cache_free_bulk(struct kmem_cache *s, size_t size, void **p)
{
  ...
  do {
    struct detached_freelist df;

    size = build_detached_freelist(s, size, p, &df);
    ...
    __slab_free(df.s, df.slab, df.freelist, df.tail, df.cnt,
           _RET_IP_);
  } while (likely(size));
}
```

In practice, the step to take is to first allocate far more than `sheaf_capacity * (MAX_FULL_SHEAVES + 1)` objects of the same cache, then free them all at once, after which the cross-cache attack can proceed through the normal flow.

## 4.3 Solution 2: Bypassing the Sheaf Cache Layer via Cross-NUMA Node Allocation

Another method is proposed here that can bypass the sheaf caching mechanism entirely and exploit the original slab mechanism directly. After an object is freed, in the free fast path, `can_free_to_pcs` decides whether the object can be cached by sheaf. If it cannot be cached, it falls back through `__slab_free` to the traditional allocation and reclamation mechanism.

mm/slub.c

```
// https://elixir.bootlin.com/linux/v7.1.4/source/mm/slub.c#L6246
static __fastpath_inline
void slab_free(struct kmem_cache *s, struct slab *slab, void *object,
         unsigned long addr)
{
  memcg_slab_free_hook(s, slab, &object, 1);
  alloc_tagging_slab_free_hook(s, slab, &object, 1);

  if (unlikely(!slab_free_hook(s, object, slab_want_init_on_free(s), false)))
    return;

  if (likely(can_free_to_pcs(slab)) && likely(free_to_pcs(s, object, true)))
    return;

  __slab_free(s, slab, object, object, 1, addr);
  stat(s, FREE_SLOWPATH);
}
```

In the original design, the sheaf mechanism would cache objects from a remote NUMA node, but caching objects from a remote NUMA node lowers the chance of hitting the local NUMA node on later allocations, which increases overhead[16]. For this reason, an object from a remote NUMA node is put directly back into its original slab.

mm/slub.c

```c
// https://elixir.bootlin.com/linux/v7.1.4/source/mm/slub.c#L6002
static __always_inline bool can_free_to_pcs(struct slab *slab)
{
    int slab_node;
    int numa_node;

    if (!IS_ENABLED(CONFIG_NUMA))
        goto check_pfmemalloc;

    slab_node = slab_nid(slab);
    ...
    numa_node = numa_mem_id();
    if (likely(slab_node == numa_node))
        goto check_pfmemalloc;
    ...
    numa_node = numa_node_id();
    /* freed object is from this cpu's node, proceed */
    if (likely(slab_node == numa_node))
        goto check_pfmemalloc;
    ...
    return false;

check_pfmemalloc:
    return likely(!slab_test_pfmemalloc(slab));
}
```

This can therefore be used when two NUMA nodes are present in the environment. If an object is allocated on the first NUMA node but freed on the second, the NUMA node of the sheaf cache layer differs from the NUMA node the object sits on, so it does not enter the sheaf mechanism but falls back to `__slab_free` to free the object at the third layer. This makes it easy to bypass the sheaf mechanism and carry out the cross-cache flow.

# 5 Sheaf Design Flaws: Inconsistency Between the Sheaf and Traditional Slab Implementations

As described in Section 3.2.1, the exploitation of the provided buffer vulnerability takes advantage of a design flaw in sheaf so that the double free detection of `CONFIG_SLAB_FREELIST_HARDENED` [17] is not triggered. Starting from this flaw, this section systematically examines how the original slab freelist's `CONFIG_SLAB_FREELIST_HARDENED` mitigation was not accounted for during development, and how its two protections, double free detection and pointer protection, are thereby rendered ineffective.

## 5.1 Flaw 1: Bypassing `CONFIG_SLAB_FREELIST_HARDENED` double free detection

In the traditional SLUB allocator, after an object is freed, its freelist pointer is updated through `set_freepointer()`. Before setting the next free object, SLUB checks whether the current object points to itself, in order to detect a double free.

mm/slub.c

```c
// https://elixir.bootlin.com/linux/v7.1.4/source/mm/slub.c#L541
static inline void set_freepointer(struct kmem_cache *s, void *object, void *fp)
{
  unsigned long freeptr_addr = (unsigned long)object + s->offset;

#ifdef CONFIG_SLAB_FREELIST_HARDENED
  BUG_ON(object == fp); /* naive detection of double free or corruption */
#endif

  freeptr_addr = (unsigned long)kasan_reset_tag((void *)freeptr_addr);
  *(freeptr_t *)freeptr_addr = freelist_ptr_encode(s, fp, freeptr_addr);
}
```

However, when storing a newly freed object, sheaf does not use `set_freepointer()`. As a result, when the same object is freed repeatedly, sheaf can store the duplicate object reference directly, and during a later allocation two structures may end up sharing the same block of memory. In addition, since the sheaf cache sits before the SLUB freelist, objects are managed by sheaf first, which lets an attacker build a usable double free primitive without triggering the SLUB freelist double free protection.

### 5.2 Flaw 2: Bypassing `CONFIG_SLAB_FREELIST_HARDENED` pointer protection

Besides double free detection, `CONFIG_SLAB_FREELIST_HARDENED` also provides freelist pointer protection, which prevents an attacker from directly modifying the next object pointer stored in the freelist. In the traditional SLUB, the next pointer of a free object is not stored in plaintext but encoded through `freelist_ptr_encode()`, which xors the next pointer with `s->random`, a random value generated when each `kmem_cache` is initialized.

mm/slub.c

```c
// https://elixir.bootlin.com/linux/v7.1.4/source/mm/slub.c#L504
static inline freeptr_t freelist_ptr_encode(const struct kmem_cache *s,
              void *ptr, unsigned long ptr_addr)
{
  unsigned long encoded;

#ifdef CONFIG_SLAB_FREELIST_HARDENED
  encoded = (unsigned long)ptr ^ s->random ^ swab(ptr_addr);
#else
  encoded = (unsigned long)ptr;
#endif
  return (freeptr_t){.v = encoded};
}
```

After the sheaf caching layer is introduced, a freed object first enters the object array managed by `slab_sheaf`. However, the sheaf object pointer is not encrypted. This renders the existing protection ineffective and lets an attacker easily hijack the pointer, thereby controlling the memory location allocated by the `kmem_cache`. How this flaw is exploited is described in detail in Section 6.

# 6 Novel Techniques: Sheaf-Based Exploitation

In the previous sections, this research built on the sheaf-related parts of the vulnerability in Section 3 to propose additional privilege escalation paths and to reduce the impact of the sheaf/barn mechanism on prior SLUB exploitation techniques. To discuss the impact of the sheaf mechanism on Linux kernel exploitation in greater depth, this section proposes three novel exploitation techniques.

Existing online discussions and articles on sheaf-related exploitation, as of July 2026, remain fairly limited. The most representative is SheafJack [18], which likewise focuses on the attack surface introduced by the sheaf mechanism and is one of the few pieces of prior work that publicly discuss this topic. However, the direction it proposes is vague in many details and contains incorrect descriptions. Taking its description of attack vector 2 as an example, it mentions triggering the cross-cache flow through `refill_sheaf`, but reviewing the actual source code shows that no such behavior exists. This section provides a more detailed and concrete explanation of sheaf-based exploitation and presents new exploitation techniques.

## 6.1 Technique 1: Object Array Hijack

`slab_sheaf` is an object pointer array. The detailed introduction of it can be found in Section 1.2. When an object is freed, the `main` sheaf is taken from the current `slub_percpu_sheaves`, the object is stored into that structure, and `size` is updated.

mm/slub.c

```
// https://elixir.bootlin.com/linux/v7.1.4/source/mm/slub.c#L5806
static __fastpath_inline
bool free_to_pcs(struct kmem_cache *s, void *object, bool allow_spin)
{
  struct slub_percpu_sheaves *pcs;
  ...
  if (unlikely(pcs->main->size == s->sheaf_capacity)) {
  ...
  }

  pcs->main->objects[pcs->main->size++] = object;
```

Conversely, when the kernel tries to allocate a block of memory, it also attempts to obtain an object from the `main` sheaf.

```
// https://elixir.bootlin.com/linux/v7.1.4/source/mm/slub.c#L4707
static __fastpath_inline
void *alloc_from_pcs(struct kmem_cache *s, gfp_t gfp, int node)
{
  struct slub_percpu_sheaves *pcs;
  bool node_requested;
  void *object;
  ...
  object = pcs->main->objects[pcs->main->size - 1];
  ...
  pcs->main->size--;

  return object;
}
```

If the object array pointer in `slab_sheaf` can be overwritten, the memory allocated by SLUB can be hijacked to any desired location for later exploitation. The `slab_sheaf` structure is shown below.

```
// https://elixir.bootlin.com/linux/v7.1.4/source/mm/slub.c#L404
struct slab_sheaf {
  union {
    struct rcu_head rcu_head;
    struct list_head barn_list;
    /* only used for prefilled sheafs */
    struct {
      unsigned int capacity;
      bool pfmemalloc;
    };
  };
  struct kmem_cache *cache;
  unsigned int size;
  int node; /* only used for rcu_sheaf */
  void *objects[];
};
```

The object array in this structure is a variable-length array of pointers. Every `kmem_cache` has its own sheaf layer, and which `kmem_cache` a `slab_sheaf` is allocated from depends on the size of the object it manages, as shown in Table 1. It is worth noting that the actual capacity is obtained by taking the maximum of the value in Table 1 and the capacity specified when each `kmem_cache` is initialized.

Table 1: Default sheaf capacity and the kmalloc cache backing `struct slab_sheaf`

| `s->size` | Capacity | Backing kmalloc cache |
|---|---|---|
| $< 256$ | 60 | kmalloc-512 |
| $\geq 256$ and $< 1024$ | 26 | kmalloc-256 |
| $\geq 1024$ and $<$ `PAGE_SIZE` | 12 | kmalloc-128 |
| $\geq$ `PAGE_SIZE` | 4 | kmalloc-64 |

```c
// https://elixir.bootlin.com/linux/v7.1.4/source/mm/slub.c#L7741
  if (s->size >= PAGE_SIZE) capacity = 4;
  else if (s->size >= 1024) capacity = 12;
  else if (s->size >= 256)  capacity = 26;
  else        capacity = 60;

  /* Increment capacity to make sheaf exactly a kmalloc size bucket */
  size = struct_size_t(struct slab_sheaf, objects, capacity);
  size = kmalloc_size_roundup(size);
  capacity = (size - struct_size_t(struct slab_sheaf, objects, 0)) / sizeof(void *);

  return max(capacity, args->sheaf_capacity);
```

Therefore, by allocating or freeing large numbers of the objects managed by `slab_sheaf`, objects in the four buckets kmalloc-64, kmalloc-128, kmalloc-256, and kmalloc-512 can be heap sprayed, and a UAF can ultimately be used to hijack the object array to achieve arbitrary read and write.

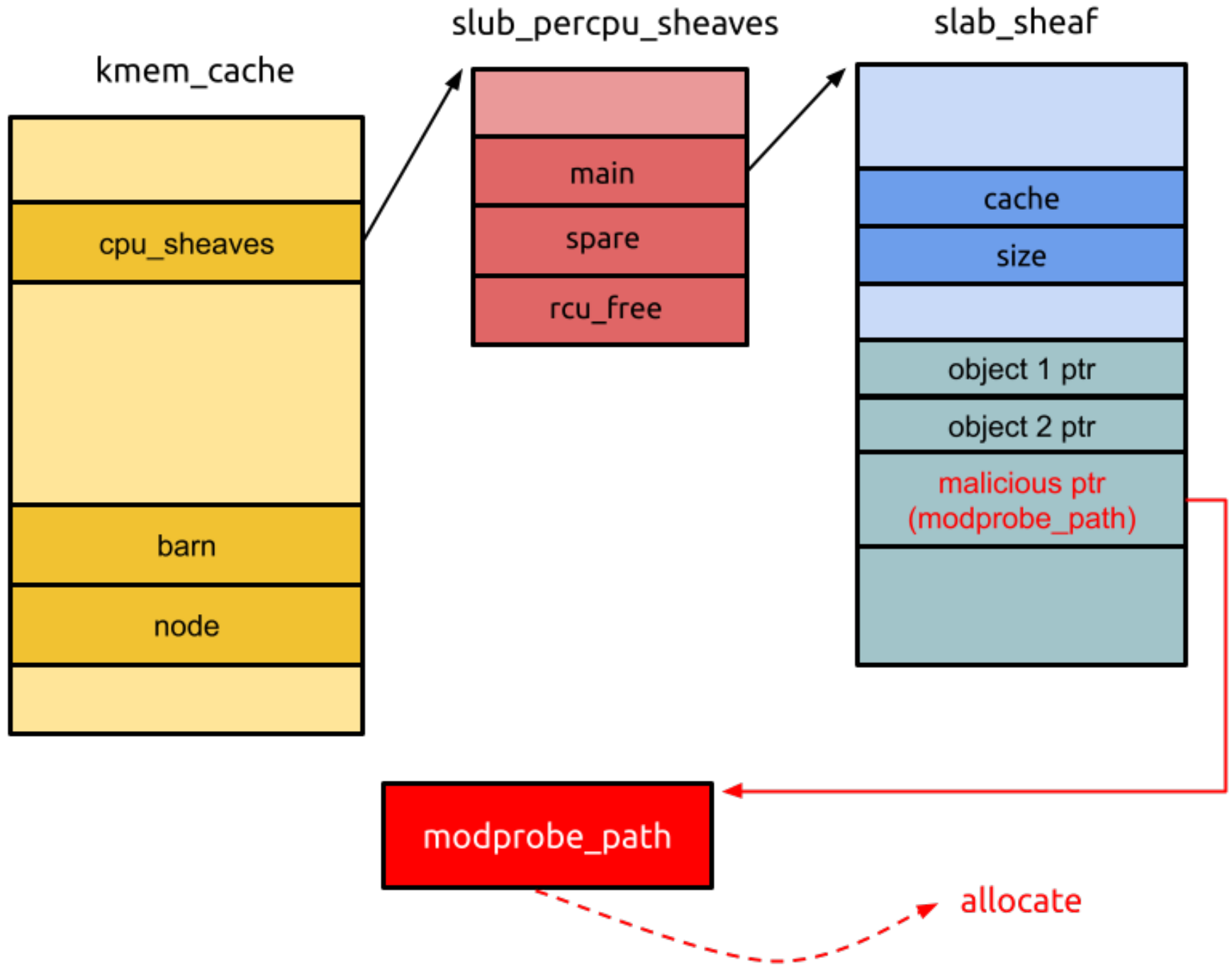

Figure 3: Technique 1, hijacking the `slab_sheaf` object array.

The actual procedure is as follows. In this technique, the goal is to reclaim the UAF object by heap spraying `slab_sheaf`. Here, a kmalloc-512 UAF is assumed to be available, and maple tree nodes can be chosen for the heap spray. Each time a new region is mmap'd, a new VMA is allocated, and a new maple node is allocated from `maple_node_cache`.

lib/maple_tree.c

```c
// https://elixir.bootlin.com/linux/v7.1.4/source/lib/maple_tree.c#L1089
static inline void mas_alloc_nodes(struct ma_state *mas, gfp_t gfp)
{
  ...
    mas->alloc = mt_alloc_one(gfp);
    if (!mas->alloc)
      goto error;
```

This kmem_cache specifies a capacity of 32 at initialization, so its slab_sheaf is allocated from kmalloc-512.

lib/maple_tree.c

```c
// https://elixir.bootlin.com/linux/v7.1.4/source/lib/maple_tree.c#L5630
void __init maple_tree_init(void)
{
  struct kmem_cache_args args = {
    .align  = sizeof(struct maple_node),
    .sheaf_capacity = 32,
  };
}
```

To heap spray slab_sheaf, the slab_sheaf instances in the barn and in slub_percpu_sheaves are first exhausted through mass allocation and freeing. Then, when allocating a new object, because all slab_sheaf instances are out of objects, the kernel will try to allocate a new slab_sheaf.

mm/slub.c

```c
// https://elixir.bootlin.com/linux/v7.1.4/source/mm/slub.c#L4649
static struct slub_percpu_sheaves *
__pcs_replace_empty_main(struct kmem_cache *s, struct slub_percpu_sheaves *pcs, gfp_t gfp)
{
  struct slab_sheaf *empty = NULL;
  ...
  if (!empty) {
    empty = alloc_empty_sheaf(s, gfp);
    if (!empty)
      return NULL;
  }
```

After that, the UAF is used to overwrite the object array with modprobe_path or any other target location. Privilege escalation is then achieved by making the allocator return memory at a target such as modprobe_path and performing an arbitrary read and write there.

> To simplify the verification process, Technique 1 and Technique 2 were verified using a custom kernel driver that contains a UAF, available at https://hackmd.io/epHcDaB_Tr6RZWmLobJY0g?view#technique-1-and-2-test-kernel-driver.

### Artifacts

- Exploit: https://hackmd.io/epHcDaB_Tr6RZWmLobJY0g?view#technique-1-exploit
- Demo: https://youtu.be/pgtzXYAjx08

## 6.2 Technique 2: Object Array Out-of-Bounds Write

An arbitrary write of sufficient length is not always available, or the UAF write primitive on the object may be incomplete, so overwriting the object array can easily lead to a kernel panic. Two additional methods are provided here to escalate privileges without controlling the object array, the first below and the second in Section 6.3. The following is the behavior when freeing an object.

mm/slub.c

```
// https://elixir.bootlin.com/linux/v7.1.4/source/mm/slub.c#L5806
static __fastpath_inline
bool free_to_pcs(struct kmem_cache *s, void *object, bool allow_spin)
{
  struct slub_percpu_sheaves *pcs;
  ...
  if (unlikely(pcs->main->size == s->sheaf_capacity)) {
  ...
  }

  pcs->main->objects[pcs->main->size++] = object;
```

It can be observed that when the `size` field of the current `main` sheaf equals `capacity`, the sheaf is treated as a full sheaf that can take no more objects, so the `main` sheaf is swapped with another sheaf. However, if the `size` field of `slab_sheaf` can be controlled to exceed `capacity`, no check is performed, and the next time an object is freed, its object pointer is written to an arbitrary location. This yields a single arbitrary pointer write.

The exploit procedure is as follows. First, two `cred` structures with root privileges are loaded into memory in advance. The `usage` field is set to a large value to prevent the structures from being accidentally reclaimed during later exploitation.

Next, `slab_sheaf` and `task_struct` are heap sprayed in an interleaved manner. `task_struct` is sprayed through `fork`, and each subprocess repeatedly checks its own uid.

Through corrupting the `size` field, the offset of the next write is changed to the `task_struct->real_cred` field. The pre-arranged structures are then freed. On the first free, the object is placed at `objects[size]` through `free_to_pcs` to overwrite the `real_cred` pointer. On the second free, it is placed at `objects[size+1]` to overwrite the `cred` pointer. After the out-of-bounds operation, the `size` is restored to its original position to prevent a kernel panic during subsequent frees.

When a looping subprocess finds that its own uid has become 0, that is root, a root shell can be opened, completing the privilege escalation.

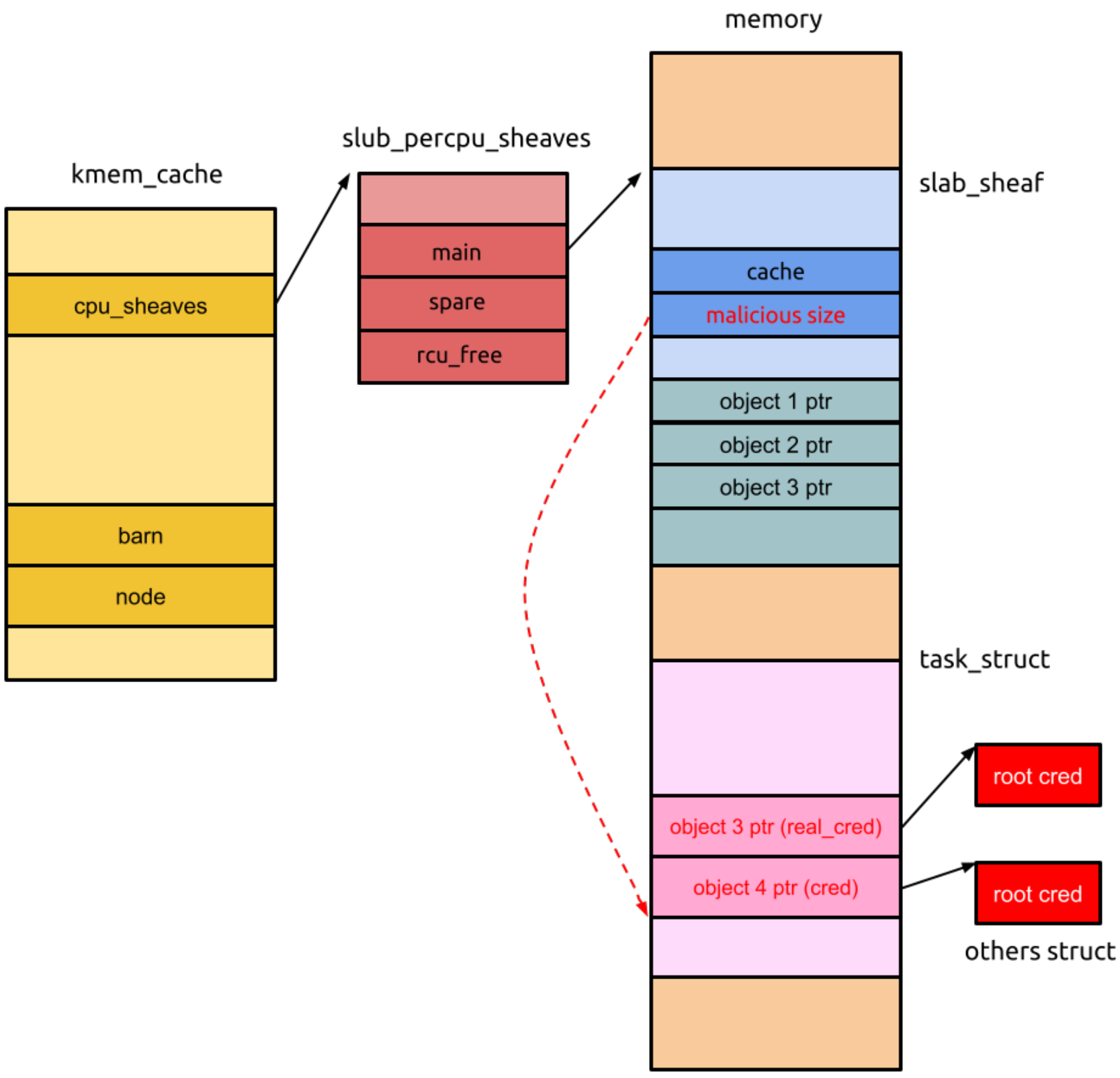

Figure 4: Technique 2, out-of-bounds write past the end of the `slab_sheaf` object array.

## Artifacts

- Exploit: https://hackmd.io/epHcDaB_Tr6RZWmLobJY0g?view#technique-2-exploit
- Demo: https://youtu.be/K4ZJ_W5pkfE

### 6.3 Technique 3: Cross-Cache via the RCU `slab_sheaf`

In this section, a novel technique is proposed that resolves the following three problems of traditional cross-cache:

- Traditional cross-cache requires mass frees to fill the SLUB partial freelist and have it reclaimed into the buddy system, and there is no precise way to know whether the page holding the UAF object has already been reclaimed.
- The operating system carries a lot of noise, so the page holding the reclaimed UAF object is very likely to be allocated by another `kmem_cache`, causing the exploitation to fail.
- When the buddy system finds that adjacent pages are all reclaimed, they are merged into a larger `free_area`, making the corresponding page difficult to obtain.

This section presents a new technique that can precisely steer a UAF object into a `kmem_cache` of any size and type, and that can escalate privileges under more constrained environments or primitives.

Besides `main` and `spare`, `slab_sheaf` also has an `rcu` slot. After an object is freed through `kfree_rcu`, the rcu sheaf stores the freed object until the grace period ends, at which point the whole batch of objects is flushed into the barn. If the barn reaches its limit, the objects are flushed directly back into the slab of the corresponding `kmem_cache`.

mm/slab_common.c

```
// https://elixir.bootlin.com/linux/v7.1.4/source/mm/slab_common.c#L1957
void kvfree_call_rcu(struct rcu_head *head, void *ptr)
{
  unsigned long flags;
  struct kfree_rcu_cpu *krcp;
  ...

  if (!IS_ENABLED(CONFIG_PREEMPT_RT) && kfree_rcu_sheaf(ptr))
    return;

  success = add_ptr_to_bulk_krc_lock(&krcp, &flags, ptr, !head);
  if (!success) {
    ...

    head->func = ptr;
    head->next = krcp->head;
    WRITE_ONCE(krcp->head, head);
    atomic_inc(&krcp->head_count);
  }
  ...
}
EXPORT_SYMBOL_GPL(kvfree_call_rcu);
```

Through the rcu sheaf, an RCU callback does not have to be created for every object, and a large number of RCU objects can be processed in batch. The callback function `rcu_free_sheaf` registered by `call_rcu` is then invoked after the grace period ends, and it begins cleaning up the objects in the rcu sheaf.

mm/slub.c

```
// https://elixir.bootlin.com/linux/v7.1.4/source/mm/slub.c#L5974
bool __kfree_rcu_sheaf(struct kmem_cache *s, void *obj)
{
  ...

  rcu_sheaf->objects[rcu_sheaf->size++] = obj;
  ...
  if (rcu_sheaf)
    call_rcu(&rcu_sheaf->rcu_head, rcu_free_sheaf);
}
```

Which barn an object is placed into, or which `kmem_cache` structure it is flushed back to, is looked up according to the `cache` field of the rcu `slab_sheaf` itself, and which `kmem_cache`'s barn is used is also determined by that field.

mm/slub.c

```c
static void rcu_free_sheaf(struct rcu_head *head)
{
  struct slab_sheaf *sheaf;
  struct node_barn *barn = NULL;
  struct kmem_cache *s;

  s = sheaf->cache;
  ...

  barn = get_barn_node(s, sheaf->node);
  ...

  if (data_race(barn->nr_full) < MAX_FULL_SHEAVES) {
    barn_put_full_sheaf(barn, sheaf);
    return;
  }
  ...
flush:
  stat(s, BARN_PUT_FAIL);
  sheaf_flush_unused(s, sheaf);
}
```

With a UAF primitive available, hijacking the `cache` field allows a large heap spray of RCU to flush the UAF object into an arbitrary `kmem_cache` during the RCU callback.

This technique is a new cross-cache path. It does not rely on the traditional route of going through the buddy system to cross a UAF object into another memory pool, reducing the uncertainty caused by internal state changes in the buddy system, and making the overall attack flow more stable. In addition, this method does not require the complex free procedure described in Section 4.2, nor the NUMA node dependent operations described in Section 4.3, which simplifies the overall execution flow and improves the portability of the technique.

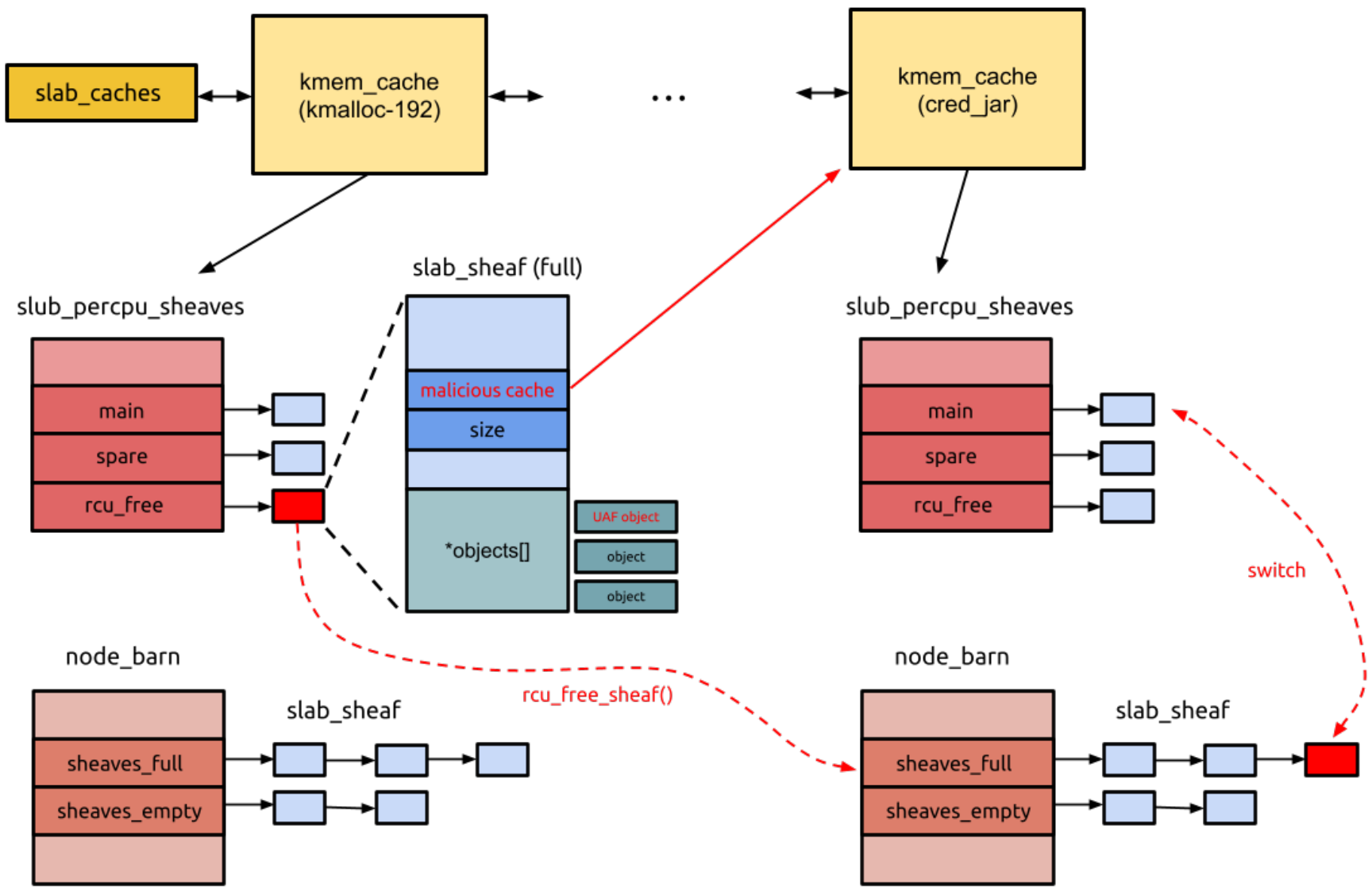

Figure 5: Technique 3, redirecting an RCU sheaf into an arbitrary `kmem_cache`.

## 6.4 Applying Technique 3 to Vulnerability 2

This section applies the proposed Technique 3 to the provided buffer vulnerability found in this research, offering a new privilege escalation path.

### 6.4.1 Triple free for information leak and overlapping the RCU `slab_sheaf`

The goal is to corrupt an object managed by kmalloc-192, whose `slab_sheaf` comes from kmalloc-512, so many of the following operations take place across these two `kmem_cache`.

First, in addition to pre-allocating `user_key_payload` and later allocating it via update key, a large number of IPv6 multicast sockets are also created in advance for the subsequent heap spray of the RCU `slab_sheaf`.

Then, the vulnerability is triggered as before, but this time a double free is performed twice, producing three pointers on the sheaf that point to the same location for later use.

After that, `user_key_payload` allocated from kmalloc-512 overlaps with the `iovec` structure, overwriting `datalen` and causing an out-of-bounds read. This out-of-bounds read is mainly used to observe the sheaf state and leak the `cred_jar` address, while the third pointer is left for later use by a newly allocated sheaf.

A large number of kmalloc-192 objects are then allocated and freed, in order to exhaust the free sheaves in the `main` and barn of the corresponding `kmem_cache`. This forces `alloc_empty_sheaf` to request new memory for the rcu `slab_sheaf` when a kmalloc-192 object is RCU freed and placed into it, and that new sheaf takes the third pointer produced above, overlapping the `iovec` with the `user_key_payload`. Whether the occupation succeeds is determined by the `READ_LEN` returned when `user_key_payload` is read, since that value is shrunk by the allocated `slab_sheaf`.

As mentioned earlier, the heap spray of the rcu sheaf slab is done through IPv6 multicast sockets. `MCAST_MSFILTER` is the API used to filter multicast sources. To support switching the multicast source filter from the old version to the new version safely, RCU is invoked, and `kfree_rcu` is called in the end.

net/ipv6/mcast.c

```c
// https://elixir.bootlin.com/linux/v7.1.4/source/net/ipv6/mcast.c#L574
int ip6_mc_msfilter(struct sock *sk, struct group_filter *gsf,
        struct sockaddr_storage *list)
{
  struct ip6_sf_socklist *psl;
  ...
  kfree_rcu(psl, rcu);

  return err;
}
```

### 6.4.2 Leaking the cred `kmem_cache` address

The `iovec` is then updated again to restore `datalen` and regain the information leak. `cred` is heap sprayed and freed through `fork` subprocesses to force the allocation of a `slab_sheaf` from the `cred kmem_cache`, so that `slab_sheaf->cache` appears in memory. The address of that structure and its fields appear near `user_key_payload`, and the location of `cred_jar` can be leaked through the information leak. Although `cred` is also kmalloc-192 in size, it is a `kmem_cache` independent of the general cache pool, so it does not affect the state produced earlier.

kernel/cred.c

```c
// https://elixir.bootlin.com/linux/v7.1.4/source/kernel/cred.c#L36
static struct kmem_cache *cred_jar;

// https://elixir.bootlin.com/linux/v7.1.4/source/kernel/cred.c#L535
void __init cred_init(void)
{
  /* allocate a slab in which we can store credentials */
  cred_jar = KMEM_CACHE(cred,
            SLAB_HWCACHE_ALIGN | SLAB_PANIC | SLAB_ACCOUNT);
}
```

### 6.4.3 Privilege escalation by poisoning the RCU cache and cross-cache attacking `cred_jar`

A UAF object is then produced again through the earlier method. One object is an `iovec` used to write to the object, and another uses an IPv6 multicast socket so `ip6_sf_socklist` can take the other copy. After RCU ends, the UAF object is placed into the RCU `slab_sheaf` that was overlapped with the other objects earlier.

Next, the RCU `slab_sheaf->cache` is written to change it to `cred_jar`.

A large heap spray is then performed so that `ip6_sf_socklist` fills the RCU `slab_sheaf`. Once it becomes a full sheaf, the `slab_sheaf` is swapped into the barn. Because its owning cache has been corrupted, it is placed into the barn of `cred`. The UAF object is thereby cross-cached into another `kmem_cache`.

Finally, through a large number of `fork` subprocesses, an attempt is made to obtain the UAF object that has entered `cred_jar`, and the UAF object is actively written through the `iovec`. Each subprocess then keeps checking in a loop whether its uid or gid has been overwritten, and privilege escalation is ultimately achieved.

### Artifacts

- Exploit: `https://hackmd.io/epHcDaB_Tr6RZWmLobJY0g?view#vulnerability-2-exploit-via-technique-3`
- Demo: `https://youtu.be/9xl1PZMr06k`

# 7 Conclusion

This research first discovers two 0-days in the io_uring subsystem. Both vulnerabilities have already been reported to upstream, fixed, and patched.

- The first vulnerability lies between the CQ/SQ ring resize feature and the deferred task work mechanism. During ring resize, the update to `ctx->rings` and the task work operation that modifies the ring flag do not hold the proper lock and therefore lack adequate synchronization, which creates a race condition that an attacker can exploit to cause a null pointer dereference.
- The second vulnerability lies in the error-handling flow of the provided-buffer bundle feature. After the kernel tries to expand the `iovec` array, an error occurs and the path rolls back when a provided buffer contains an illegal address. On the rollback path, the `iovec` pointer in `kmsg` itself is not updated correctly, leading to a UAF. This research further builds a complete exploit chain for this vulnerability and achieves local privilege escalation in an environment with multiple modern Linux kernel protections enabled.

However, during exploit development, this research finds that the SLUB sheaf/barn mechanism introduced after Linux kernel 6.18, which adds two cache layers targeting per-CPU and NUMA on top of the original slab mechanism, significantly changes the object management flow of the traditional SLUB allocator. The techniques widely used in past kernel exploitation are invalid once sheaf is introduced. For instance, once sheaf/barn introduces the extra per-CPU and NUMA node level caches, cross-cache attacks are affected: a freed object may remain in the new caching layer for a long time, so the object is marked as still in use by the slab itself and cannot be reclaimed by the buddy system. Two solutions are therefore proposed:

- More objects are heap sprayed to swap large numbers of per-CPU sheaves into the barn. Once the full sheaves or empty sheaves in the barn reach their limit, they are flushed into the slab, and the original cross-cache can finally be carried out this way.
- In an environment with multiple NUMA nodes, memory can be allocated by binding the current process to one NUMA node while freeing on another. Since caching objects from a different NUMA node in the sheaf layer would lower the chance of hitting the local NUMA node later and cause a performance problem, sheaf does not cache objects from a remote NUMA node, and this allows the original slab mechanism to be exploited directly.

In addition, a deeper study of the sheaf/barn mechanism reveals that its design renders the original SLUB `SLAB_FREELIST_HARDENED` mitigation almost completely ineffective. This mitigation mainly provides the following two protections:

- Double free check: when an object is linked into the freelist, whether the object being linked already exists in the freelist is checked, in order to detect repeated frees. On the sheaf fast path, however, this check is never used, and under normal conditions, an object is cached by sheaf/barn first, so a simple double free of the same object into the sheaf makes later exploitation easy.
- Pointer protection: the freelist is a linked list in which each freed object points to the next through a pointer, and this protection encodes the pointer by xoring it with a random value generated when the `kmem_cache` is created. In the sheaf mechanism, however, `slab_sheaf` uses an object pointer array to manage the objects stored in it, and on free or allocation, an object is obtained according to the number of objects the current sheaf holds(`size`). The object pointer array is not protected by pointer encryption, so a simple heap spray of `slab_sheaf` and tampering with the object array pointer makes it easy to hijack object allocation and freeing.

Beyond that, introducing sheaf/barn does not only bring inconvenience to SLUB exploitation. On the contrary, this mechanism actually offers a larger attack surface, and building on it, three new sheaf-based exploitation techniques are developed:

- object array hijack: this exploits the fact that `slab_sheaf->objects[]` is not covered by pointer protection to directly control the SLUB allocator into returning an object at an arbitrary location.
- object array oob write: this uses controllable sheaf metadata, such as the `size` field, to cause an out-of-bounds write on the object pointer array. Heap grooming arranges `task_struct` and `slab_sheaf` into a specific layout in memory, and tampering with the `size` then achieves a single arbitrary pointer write. The memory that this pointer refers to is filled with a root-privileged `cred`, and the arbitrary pointer write is used to overwrite `task_struct->cred` for privilege escalation.

- cross-cache by RCU sheaf: traditional cross-cache relies on the buddy system to reclaim a useless page (slab) once the partial freelist reaches its limit, thereby moving a UAF object into another `kmem_cache`. This technique instead makes use of how RCU accumulates rcu objects into a `slab_sheaf` during a grace period and creates a callback for multiple objects at once. After the grace period ends, the objects stored in the `slab_sheaf` are flushed back into the barn or the slab. `slab_sheaf->cache` records which cache the stored objects come from, and hijacking this metadata makes it possible to place objects into any cache pool more flexibly and stably. Compared with traditional cross-cache, this method does not depend on the buddy system reclaiming pages, avoiding the uncertainty caused by the buddy system's internal state and by different `kmem_cache` contending for pages.

Finally, based on the new cross-cache technique proposed above, a new exploit chain is built for the second vulnerability.

In summary, this research not only presents two new memory safety vulnerabilities in the io_uring subsystem but also builds a complete privilege escalation path on one of them. This also serves as the starting point for an in-depth study of how the new SLUB sheaf/barn allocator affects existing kernel exploitation techniques, along with the corresponding solutions. Going further, this research analyzes the sheaf/barn mechanism as a new attack surface, first showing how its design flaws render the original SLUB mitigation ineffective, and then proposing three new exploitation techniques based on the mechanism. Among them, the RCU sheaf-based cross-cache technique breaks the traditional cross-cache attack's dependence on buddy system page reclaim, allowing an attacker to steer objects between different cache pools more flexibly and stably. This research is intended to provide a new direction and attack surface for future work on Linux kernel exploitation.

## References

[1] Jens Axboe. *Faster IO through io_uring*. Kernel Recipes 2019. 2019. URL: https://www.slideshare.net/slideshow/kernel-recipes-2019-faster-io-through-iouring/176705140 (visited on 08/10/2026).

[2] frevib. *io_uring echo server benchmarks*. 2020. URL: https://github.com/frevib/io_uring-echo-server/blob/io-uring-feat-fast-poll/benchmarks/benchmarks.md (visited on 08/10/2026).

[3] Diego Didona et al. "Understanding Modern Storage APIs: A Systematic Study of libaio, SPDK, and io_uring". In: *Proceedings of the 15th ACM International Systems and Storage Conference (SYSTOR '22)*. Haifa, Israel: ACM, 2022, pp. 120–127. DOI: 10.1145/3534056.3534945. URL: https://atlarge-research.com/pdfs/2022-systor-apis.pdf (visited on 08/10/2026).

[4] Vlastimil Babka. *[PATCH v8 00/23] SLUB percpu sheaves*. Linux Kernel Mailing List. Sept. 10, 2025. URL: https://lists.openwall.net/linux-kernel/2025/09/10/485 (visited on 08/10/2026).

[5] Vlastimil Babka. *[PATCH RFC v2 00/20] slab: replace cpu (partial) slabs with sheaves*. Linux Kernel Mailing List. Jan. 12, 2026. URL: https://patchew.org/linux/20260112-sheaves-for-all-v2-0-98225cfb50cf@suse.cz/ (visited on 08/10/2026).

[6] *io_uring_setup(2) — Linux manual page.* URL: https://man7.org/linux/man-pages/man2/io_uring_setup.2.html (visited on 08/10/2026).

[7] *io_uring_enter(2) — Linux manual page.* URL: https://man7.org/linux/man-pages/man2/io_uring_enter.2.html (visited on 08/10/2026).

[8] *io_uring_register(2) — Linux manual page.* URL: https://man7.org/linux/man-pages/man2/io_uring_register.2.html (visited on 08/10/2026).

[9] Waiman Long. *mm: memcg/slab: create a new set of kmalloc-cg-<n> caches.* Linux kernel commit 494c1dfe855e. 2021. URL: https://github.com/torvalds/linux/commit/494c1dfe855ec1f70f89552fce5eadf4a1717552 (visited on 08/10/2026).

[10] *io_uring_sqe_set_flags(3) — Linux manual page.* URL: https://man7.org/linux/man-pages/man3/io_uring_sqe_set_flags.3.html (visited on 08/10/2026).

[11] *io_uring_prep_recv(3) — Linux manual page.* URL: https://man7.org/linux/man-pages/man3/io_uring_prep_recv.3.html (visited on 08/10/2026).

[12] *io_uring_setup_flags(7) — Linux manual page.* URL: https://man7.org/linux/man-pages/man7/io_uring_setup_flags.7.html (visited on 08/10/2026).

[13] Le Wu. *Game of Cross Cache: Let's Win It in a More Effective Way!* Black Hat Asia 2024. Baidu Security, 2024. URL: https://i.blackhat.com/Asia-24/Presentations/Asia-24-Wu-Game-of-Cross-Cache.pdf (visited on 08/12/2026).

[14] Dong-ok Kim, Juhyun Song, and Insu Yun. "CROSS-X: Generalized and Stable Cross-Cache Attack on the Linux Kernel". In: *Proceedings of the 2025 ACM SIGSAC Conference on Computer and Communications Security (CCS '25).* Taipei: ACM, Oct. 2025, pp. 216–230. ISBN: 979-8-4007-1525-9. DOI: 10.1145/3719027.3765152. URL: https://kaist-hacking.github.io/pubs/2025/kim:crossx.pdf (visited on 08/12/2026).

[15] Lukas Maar et al. "SLUBStick: Arbitrary Memory Writes through Practical Software Cross-Cache Attacks within the Linux Kernel". In: *33rd USENIX Security Symposium (USENIX Security 24).* Philadelphia, PA: USENIX Association, Aug. 2024, pp. 4051–4068. ISBN: 978-1-939133-44-1. URL: https://www.usenix.org/conference/usenixsecurity24/presentation/maar-slubstick (visited on 08/12/2026).

[16] Vlastimil Babka. *[PATCH v6 06/10] slab: skip percpu sheaves for remote object freeing.* maple-tree mailing list. Aug. 27, 2025. URL: https://lists.infradead.org/pipermail/maple-tree/2025-August/004477.html (visited on 08/12/2026).

[17] Kees Cook. *[PATCH v3] mm: Add SLUB free list pointer obfuscation.* kernel-hardening mailing list. July 6, 2017. URL: https://www.openwall.com/lists/kernel-hardening/2017/07/06/1 (visited on 08/10/2026).

[18] Antonius. *SheafJack: Advanced Kernel Exploitation in Linux 6.18+ via slab_sheaf Hijack.* Blue Dragon Security Research Lab. 2026. URL: https://www.slideshare.net/slideshow/sheafjack-advanced-kernel-exploitation-in-linux-6-18-via-slab_sheaf-hijack/287245786 (visited on 08/10/2026).